# A New Sentinel Approach for Energy Efficient and Hole Aware Wireless Sensor Networks

Dame DIONGUE

Ph.D Student, Department of Computer Science
Gaston Berger University
Saint Louis, Senegal
ddiongue.ep2112812@ugb.edu.sn

Ousmane THIARE

Department of Computer Science
Gaston Berger University
Saint Louis, Senegal
ousmane.thiare@ugb.edu.sn

*Abstract*—**Recent advances in micro-sensor and communication technology have enabled the emergence of a new technology, Wireless Sensor Networks (WSN). WSN have emerging recently as a key solution to monitor remote or hostile environments and concern a wide range of applications. These networks are faced with many challenges such as energy efficiency usage, topology maintenance, network lifetime maximization, etc. Experience shows that sensing and communications tasks consume energy, therefore judicious power management can effectively extend network lifetime. Moreover, the low cost of sensor devices will allows deployment of huge number nodes that can permit a high redundancy degree. In this paper, we focus on the problem of energy efficiency and topology maintenance in a densely deployed network context. Hence we propose an energy aware sleep scheduling and rapid topology healing scheme for long life wireless sensor networks. Our scheme is a strong node scheduling based mechanism for lifetime maximization in wireless sensor networks and has a fast maintenance method to cover nodes failure. Our sentinel scheme is based on a probabilistic model which provides a distributed sleep scheduling and topology control algorithm. Simulations and experimental results are presented to verify our approach and the performance of our mechanism.**

*Keywords-component; energy conservation; lifetime maximization; topology maintenance; insert (key words)*

## I. INTRODUCTION

Recent technological advances in microelectronics have favored the development of tiny and intelligent embedded devices called sensor nodes that can detect and send relevant information relatively to a given environment. This has led to the emergence of a new technology, Wireless Sensor Networks. A typical Wireless Sensor Network consists of a huge number of tiny sensor with sensing, processing and transmission capabilities [1]. These last decades, wireless sensor technology holds the lead of the stage in several sectors such as environmental monitoring, military surveillance [2], medical diagnosis [3][4], building automation [5][6], industrial automation tasks, etc. In most cases, the area of interest (wireless sensor network's deployment area) is harsh or even impossible to access for human intervention. Therefore, the deployment is most often done by airplane dropping and this may lead to unfair repartition of sensor nodes through the monitored region.

Beside problems related to random deployment, Wireless Sensor Networks are also suffering to many challenges such as data aggregation, routing, security, energy management, topology management, etc. The two later issues are attracted more and more interest from researchers and are typically addressed in this paper. Energy consumption and topology changes are of critical importance regarding Wireless Sensor Networks because the sensor node lifetime is closely related to its battery power and once deployed, they are usually inaccessible to be replaced nor recharged, due to harsh environment. However, the protocol designers should take into consideration these constraints and allow sensor nodes to have sufficient autonomy to organize themselves and cooperate with each other to save their energy. In some types of applications, random deployment is most often used and it does not always guarantee better coverage and rational use of energy. This type of deployment, may issue to energy or coverage holes problems due to unfair repartition of sensor nodes.

In this paper, we focus on node scheduling and propose an energy aware sleep scheduling and fast topology maintenance algorithm for lifetime maximization in wireless sensor networks. Our scheme is based on the Sentinel concept and need to operate in highly dense networks. The proposed scheme consists of two parts, the sleep scheduling procedure that uses nodes redundancy and dynamic probe rate adjustment to take better advantage of the redundancy, and the fast recovery procedure to take into account nodes failure. Since Sentinel scheme operate in very dense networks, it must be coupled with an effective recovery procedure. So that, if a sentinel node fails, whether in the shortest time a spare to take over and maintains the hole. Unlike the scheme proposed in [10] where authors assume an active messaging status for active nodes, here we propose that working nodes use passive messaging to limit the overhead charge. Another major challenge in wireless environment is the problem of collisions. In some case, collision may occur and cause activation of multiple nodes in a single area. To solve this problem, we use a disabling procedure, called activity withdrawal algorithm, between active nodes based on proximity and activity duration weight. When we have two active conflicting nodes, the disabling procedure permit to select the older one to ensure the monitoring task and put the other node in sleep mode.

The remainder of this is organized as fellow. Section II describes some related works in the literature. Section III details our model description and the scheduling problem definition. Section IV makes an overview to the proposed scheme. Section V shows the simulations and experimental results. Finally, section VI provides conclusion and future works.

## II. RELATED WORK

### A. Energy conservation

Wireless sensor devices are very constrained in term of battery power. Sensor nodes are non rechargeable battery operating devices and generally deployed in often inaccessible environment like forests for fire or pollution detection, sea for tracking some species, battlefields for enemies tracking, etc. Then, the only way to keep alive the network for longer time is to efficiently manage the battery power usage. However, many mechanisms, algorithms and protocols have been proposed in routing, clustering, data agregation, security, mobility, and especially coverage and connectivity areas.

Virmani and al. propose an energy efficient data agregation protocol based on nodes clustering [7]. Their protocol relies on the reduction of the distance between communication nodes. In the same vein, Murthy and al. Proposed a crosslayered clustering protocol [8]. We find that most of the works on lifetime maximization deal at the same time with the coverage problems. In [9][10], the authors use the distribution of the interest area into several cover sets (disjoint and/or nondisjoint) to efficiently rationalize the energy usage.

Other works focus on lifetime maximization based on energy efficient coverage and state management mechanisms. Achieving this assumes that nodes cooperate with each other to make distributed decisions on the choice of active subset; hence the need to synchronize the whole network activities [11][12]. This approach requires some processing and communication cost at each node. However, it is more scalable and more flexible for nodes failures. Ye and al. proposed in [12] a probing environment and adaptive sensing mechanism. They assume to activate the minimum set of nodes, over a highdensity sensor network, that can provides the monitoring of the interest area and put all the redundant nodes in sleep mode. In PEAS [12], authors proposed energy conservation by maintaining all working nodes by a minimum distance c. The asleep nodes may wake up after a random period and check their vicinity (for a radius c) by sending broadcast messages. They will enter on-duty mode only if they receive no replies from working nodes; otherwise they will stay on off-duty mode. Their solution offered a crucial benefit in term of energy consumption and guarantee an asymptotic network connectivity. But authors assume that working nodes never go back to sleep, which may result in redundant working nodes when collisions occur at the probe requesting/replying steps.

### B. Topology Maintenance

A Wireless Sensor Network well-functioning strongly depends on: (i) a good coverage of the interest area to retrieve relevant information, (ii) a good connectivity between sensor nodes to better relay information toward the Sink node, (iii) and also a good energy management policy for a long life network. However, the deployment strategies (deterministic or random) have a great influence on above criteria. Ideally, a deterministic deployment is desirable, but in most cases the monitored region, for example battlefield, is difficult or dangerously accessible and thus, a random deployment remains the only possible alternative. This deployment method often leads collateral problems such as sparse or not at all covered areas. Several solutions has been proposed in the literature in order to solve the related problems to the network topology changes. And these solutions can be classified according three approaches: node adaptation, link adaptation and mobility (mobile sensor node or robot) [13]. Node adaptation techniques are often based on: (i) clustering which propose the network to have an hierarchical organization, (ii) set cover computation which organize the network into multiple subset where each one can cover the whole network for a period of time, (iii) and lastly node scheduling technics that relies on deploying redundant nodes and schedule their activity. Gupta and al. [14] use a node scheduling technique for topology healing and a probabilistic approach to determine the coverage redundancy degree. They schedule nodes activities on the one hand to save energy and also ensure a better coverage. Always in the same direction, Corke and al. propose in [15] two algorithms. The first algorithm uses neighbors informations to detect failed nodes and determine hole location. The second algorithm uses routing informations to detect a hole from a distance and try to maintain the routing path. Their solution require that nodes keep state informations into memory. Other solution [16][17] in the literature use another approach for topology healing, link adaptation technic and this consist of adapting communication parameters and exchanging neighbors informations. Others use mobility [18] to solve the holes problems related to coverage/energy. Works in [19][20][21], opt for an additional deployment of mobile nodes (generally robots with GPS) to maintain the coverage. These solutions offer effective holes healing but generate a high network load added to that gluttony in energy of the GPS module.

In this paper, we opted for a scheduling based solution rather than deploying additional mobile nodes. Because, energy should be well tuned in Wireless Sensor Networks. However, mobility based solutions, in addition to the expensive costs of equipments, use GPS, which is very energy intensive. And also, mobility is often not easy or not at all applicable to some regions because of their relief.

## III. MODEL AND PROBLEM DESCRIPTION

### A. Network model and problem description

We first present in this section some keys definitions and properties related to our proposed algorithm.

### Definition 1: Sentinel Network Design

Here we assume a flat network with a huge number of sensor node uniformly deployed in an interest area (a network with a high density of sensor nodes). And all nodes are initially in sleep mode for a while. When wakes up, node probe their vicinity looking for a sentinel (an active node that stands in guard for a dedicated area). If the probing operation is positive i.e. a sentinel node responds by sending a probe reply message, it turn back to sleep mode else it starts the guard round.

### Definition 2 : Redundant Node Sleep Scheduling

We consider a sensor network with a huge number of nodes uniformly deployed in an interest area. The concept Redundant Nodes Sleep Scheduling (RNSS) consist of putting on off-duty all the redundant nodes and just let a minimum set of sentinel nodes that can ensure the require monitoring. In [22], authors explain the concept of completely redundant node. Therefore, according to figure 1, node $n_2$ (node $n_2$ 's communication range is represented with dashed line) will be on off-duty mode because its area is covered by nodes $s_1$; $s_5$; $s_6$ and $s_{10}$. So $n_2$ can now direct itself to sleep for $t_s$ seconds.

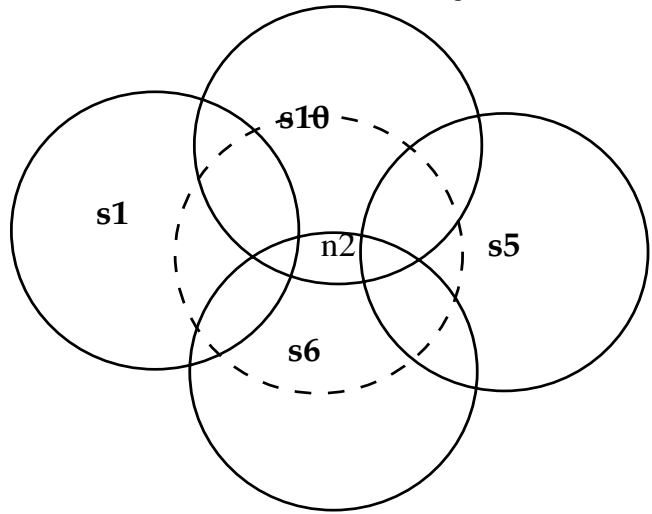

Figure 1.  Complete coverage redundancy

### Definition 3 : Strongly Connected Active Nodes

In our designed scheme, two active nodes must keep far from each other with a distance threshold $\delta = 2Rs$. Rs represent a node sensing radius since it consist of a unit disk. And we consider the communication radius of each is $R_c$ with $Rc \geq Rs$. This will permit to design a network in which all active nodes share better connectivity between them. We control nodes connectivity by adjusting the distance threshold $\delta$, then this will permit us to explore better coverage degree with the scheme.

Our aims on this paper are to: (i) minimize the subset of sentinel nodes (on-duty nodes) use to monitor the interest area; (ii) minimize the energy usage at each sentinel sensor node; (iii) and finally design a fast topology recovery procedure. To do this, we will deem the deployment of a dense network (like in Definition 1) creating a high redundancy. Thus we propose to exploit that redundancy by activating the minimum subset of sensor nodes for the monitoring and put the rest in reserve and then give sufficient autonomy to reserved nodes to distributely manage their sleep time and adjust it with the probability of failure.

### B.  Sentinel Scheduling Problem

The random deployment often causes an unbalanced distribution of nodes through the monitored region. Then, if an active node fails by battery depletion or anything else, the area which was covered by that node will remains unmonitored (creation of coverage hole). And so, all the events that occur there, will pass unnoticed. Hence, to solve this problem, we consider $\Omega$, the population of sensor nodes uniformly deployed in the interest area. And nodes have sufficient autonomy to organize themselves and select a minimum subset S where $S \subset \Omega$ of sentinel nodes. Hence, the subset $\bar{S}$ defined by $\bar{S} = \Omega - S$ falls into off-duty mode to conserve the energy. And finally nodes execute an algorithm that stands on a probabilistic scheme to control the off-duty nodes' wakeup.

### IV.   SENTINEL SCHEME

### A.  Node state transition

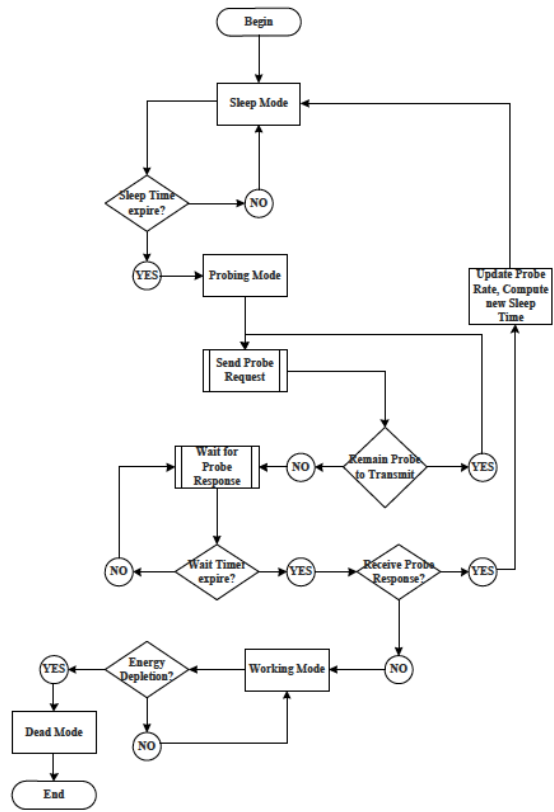

Figure 2.   Sentinel node's state transition algorithm

We consider that a sentinel node can be in one of the four following states: sleeping, probing, working or dead.
**Sleep mode:** The sleep mode corresponding to node's initial state, where they turn off their radio module. We chose to turn off at sleep mode only the radio module, because it is difficult or impossible to put a node completely off-duty.

**Probing mode:** The second mode in which a sentinel node can be is the probing mode where it has the ability to send/receive only control messages to/from its neighbors. From that state, a sentinel node can be active either go back to sleep mode.

**Active mode:** A node goes into active mode if and only if it detects no active neighbor. However, it starts to fulfill its role of sentinel node, that is say, to continuously monitor its dedicated area until the exhaustion of its battery or the reception of a probe reply from an older sentinel node (see activity withdrawal algorithm). Thus, the possible state transition from this mode is either going back to sleep or the death of the sensor.

**Dead mode:** Finally, the dead mode often characterizes sensor node's energy depletion (total battery exhaustion). This may also be due to a dysfunction of hardware component like sensing unit, communication module, etc.

### B. Sleep Scheduling Algorithm

For a suitable energy usage, we designed a scheduling algorithm that permit to select just a minimum set S of onduty nodes to ensure the monitoring task. Then all the other nodes (redundant nodes) will be left on off-duty mode (Sleep mode) that is say, they turn off their radio module. The onduty subset selection (sentinel nodes selection) is done by the following rule "the first nodes that wakes up and find no other one in its vicinity, stands guard i.e. stands as sentinel node". Hence, all the nodes are initially deployed in sleep mode and each sensor node must be asleep for $t_{s\_initial}$. After the $t_{s\_initial}$ timer expire, node should probe their vicinity by sending probe request messages to look for an active neighbor. After several series of tests, we then sets the probe reply wait timer, $t_w$, to 1 second. After the $t_w$ timer expire with no replies received, it immediately enter in active mode to monitor its vicinity. In case the probing node receives a reply from its neighbors, it check first if the responder is not far away from the distance threshold in respect to *SCAN* property. If *SCAN* is verified, node update its probe rate according to theorem 2 and then computes its new sleep time (theorem 1). Else, the node ignore the message and start its activity.

### **Theorem 1:** The Sleep Time Computation

The wake-up timer of a given node is computed with a distributed scheme using the Weibull distribution probability and node's probe rate. The Weibull distribution is most suitable for our design because it permit to adjust node's sleep time when needed. Experience shows that electronic devices failure rate grows over time and therefore, the Weibull distribution will permit to compute decreasing sleep time over simulation or over network's operating time. The Weibull parameters i.e. scale parameter $\lambda$ and the shape parameter $\beta$ are chosen as follows: the Weibull scale parameter represents node's probe rate and is a function of time while the shape parameter is a value selected from $\{1.5, 2.0, 3.0\}$. We started the shape parameter's values at 1.5 because if $\beta = 1.0$, we have an exponential distribution.

***Proof :*** We suppose that $R$, uniformly generated in range

$[0,1[$, is the probability of awakening of a given node denoted by $X$ and obtained by the Weibull cumulative distribution function. We aim to determine $t$ such that :

$$R = 1 - F(t) = 1 - P[X \le t] = 1 - \int_0^t f(u)du$$

Therefore, we have :

$$R = 1 - \left(1 - e^{-\left(\frac{t}{\alpha}\right)^\beta}\right)$$

$$\Rightarrow R = e^{-\left(\frac{t}{\alpha}\right)^\beta} \qquad (1)$$

And then, we deduce from equation (1) a node's sleep time $t_s$ by applying the logarithm :

$$lnR = \ln\left(e^{-\left(\frac{t}{\alpha}\right)^\beta}\right) = -\left(\frac{t}{\alpha}\right)^\beta$$

$$t_s = \alpha \ln\left(\frac{1}{R}\right)^{1/\beta} \qquad (2)$$

Where $\alpha = 1/\lambda$ and $\beta$ are respectively the Weibull scale and shape parameters. We define that $\lambda$ represents a node's probe rate [12].∎

Another significative contribution presented in this paper is that nodes are sufficiently autonomous for updating their probe rate used to calculate the sleep time. For this, we use the Weibull hazard function $h(t)$. And unlike PEAS and LDAS, here nodes have no need to keep neighbor information for the scheduling procedure. Some solutions in the literature use neighbors informations to take some decision. However, we propose to dynamically update nodes probe rate and this is done independently from neighbors informations (refer to Theorem 2). Thus our scheme scheme is designed to be completely distributed.

### **Theorem 2 :** Dynamic probe rate adjustment

Before a sleep round, each node must compute its new probe rate based on the network's lifetime and its old probe rate. This is done at each node independently from its neighbor.
***Proof :*** Using the Weibull hazard function we obtain from the survival function, we have :

$$h(t) = \lim_{\Delta t \to 0} \frac{1}{\Delta t} P(t < X \le t + \Delta t | X > t)$$

$$h(t) = \frac{f(t)}{1 - F(t)}$$

Then, we have :

$$h(t) = \frac{\beta}{\alpha}\left(\frac{t}{\alpha}\right)^{\beta-1}$$

And $h(t)$ represents the new probe rate ( $h(t) = \lambda(t)$ ).∎

**Input:** $s_i, s_j$: two active nodes;
$d(s_i, s_j)$: distance between nodes $s_i$ and $s_j$;
$a.s_i, a.s_j$: activity duration of node $s_i$ respectively node $s_j$;
$\delta$: distance threshold between two active nodes;

**initialization:** nodes receive probe replies;

**if** $d(s_i, s_j) \leq \delta$ **then**
**break** ;
**else**
**if** $a.s_i \leq a.s_j$ **then**
Node $s_i$ computes a new sleep timer $t_s$;
$a.s_i \leftarrow 0$;
Node $s_i$ turns back to sleep mode for $t_s$;
**else**
Node $s_i$ ignore the received reply;
**end**
**end**

Algorithm 1 : Activity withdrawal algorithm

### C. Activity withdrawal algorithm

Due to collisions, active nodes may be conflicting. And to solve this problem, we introduce an activity withdrawal procedure Algorithm-1. When they receive probes request messages, sentinel nodes may response by sending a probe reply message. Probe replies include the sender's coordinates *(x; y; z)* and it activity age *a.si*. In case where there are two or more conflicting active nodes, they all execute the withdrawal algorithm. Let us consider the scenario in Fig.3. Let us consider in this example a domain with the nodes n0, n4, n5, n10 and n20 all initially in sleep mode. They have all sleep timer randomly generated according to the Weibull distribution. Thus, nodes can wake up at different dates. Suppose that node n5 wakes up first and finds that there is no active neighbors, so it immediately starts its activity (n5 → s5). After a while, the other nodes can wake up and scan the vicinity by sending probes request messages. The sentinel node s5 by receiving probes request, will responds to its neighbors. His response may include its position and its activity duration (elapsed time since the beginning of its activity until the response). Nodes that hear node s5 's response will check first if the *SCAN* property is verified. If yes, they update their probe rate and then compute a new sleep timer; otherwise they ignore the message. Since probes messages are broadcasted and the node's wake up are not synchronized, it is possible that collision occur and thereby prevent some messages reaching their destination. The node n0 will wait until its tw expires to start its activity. Thereby, we will have two sentinel nodes in the same area. Thus, to solve this problem, we introduce the activity withdrawal algorithm (algorithm 1) that permit to disable the youngest sentinel node.

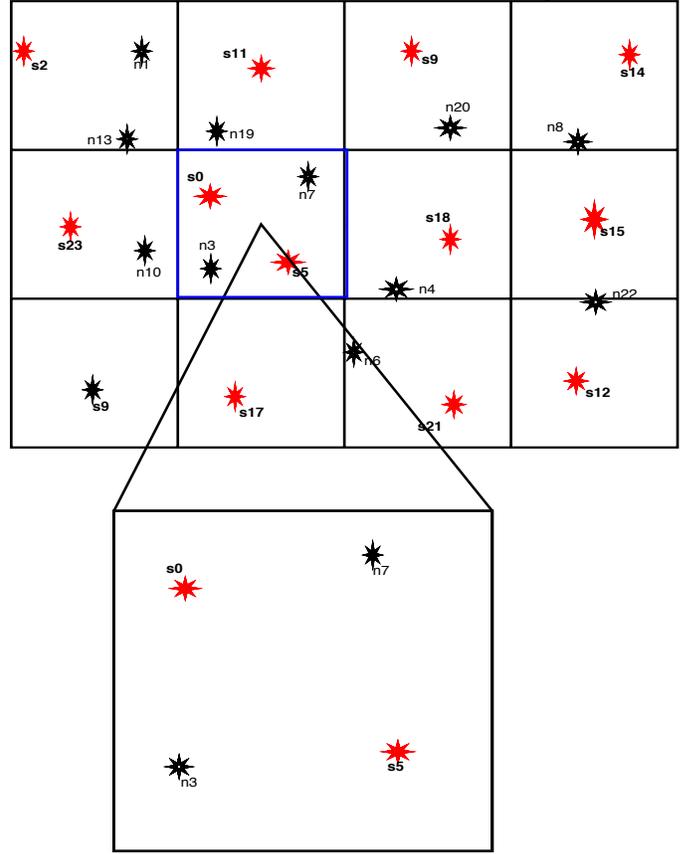

Figure 3.   An example to illustrate redundant active node scenario

### D. Topology healing procedure

As detailed in Fig. 2, after deployment, a subset of sentinel nodes monitor the region of interest until their energy exhaustion. When a sentinel node fails, one of its neighbors in the reserve subset will wakes up to fill the leaved hole. For more clarity, let us consider the scenario in fig. 4. Initially we have S = {S0; S1; S8; S13; S15; S20} and these nodes monitor the interest region until energy depletion. After a while, sentinel nodes S0 and S8 fail and the sentinel subset becomes S = {S1; S4; S7; S12; S13; S15; S20}. Looking at this subset S, we find that there are four new sentinel nodes that is say S4; S7; and S12 . Without such redundancy, coverage holes can be created with the loss of nodes over time. As in the example of the Fig. 5, S2 and S4 die and leave uncovered their dedicated area. Since there is no node in reserve to compensate for the vacuum, the only alternative is the deployment of mobile nodes that requires a GPS guidance.

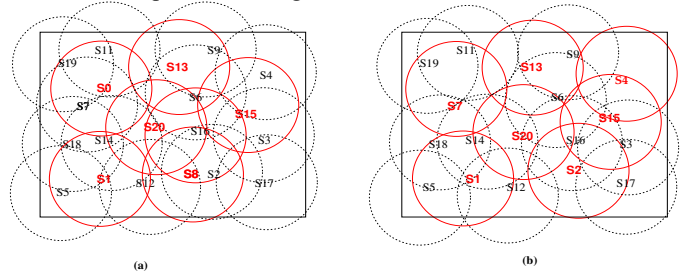

Figure 4.   Coverage hole maintenance in redundant deployment scenario

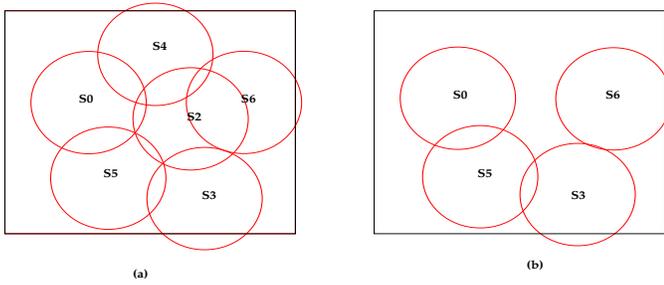

Figure 5.   Coverage hole creation with non redundant deployment scenario

## V.   PERFORMANCE EVALUATION

In this section, we will evaluate our algorithm by measuring the control overhead charge and by comparing it with other algorithm in term of energy usage efficiency. To show that our algorithm is energy efficient, it will be compared to PEAS algorithm. Our scheme will be compared to that in [12] with performance ratio.

### A.   Simulation model and parameters

We have built a distributed node scheduling algorithm to perform network lifetime in wireless sensor networks. We simulate our scheme using Castalia [1], an OMNeT++ [2] framework designed for wireless sensor networks. For experimentations, we deployed uniformly the sensor nodes in a flat network. Sensor nodes are *2AA* battery equipped and are randomly deployed, initially in sleep mode, in a square field of *50 meters x 50 meters*. To be close to the reality, we assume that channel condition is not perfect and nodes sensing range is defined to *10 meters* ($\delta \leq 2R$ *ie* $\delta \leq 20$ *meters*). So that the probability that collision occur is not zero. Then to avoid much overhead processing, we choose small control messages (*25 octets* by default) to ensure nodes communications.

### B.   Energy efficiency evaluation

Fig. 6 shows an evaluation of the average energy consumption with different values of $\beta$ parameter. And we can see that varying the $\beta$ parameter has no major influence on the energy consumption so on the network lifetime.

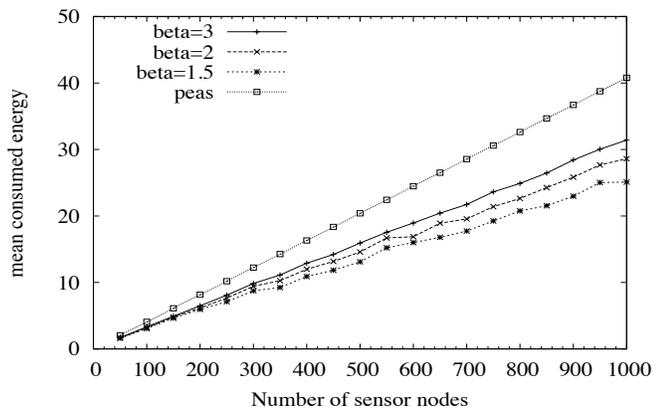

Figure 6.   Average energy consumption by varying $\beta$ parameter

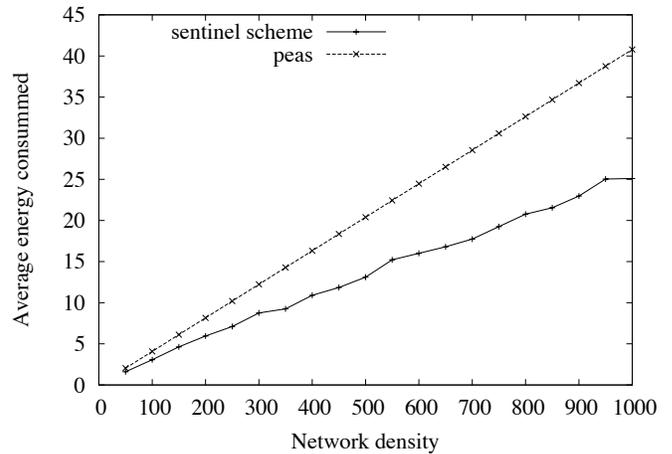

Figure 7.   Average energy consumption between Sentinel scheme and PEAS

We choose to vary that parameter because it permit a generalization of some other probabilistic distributions such as Exponential ($\beta = 1$) or Rayleigh distribution ($\alpha = 1$, $\beta = 2$). We simulated the networks for 6000 seconds and we measured the average energy spent by the whole network and finally compared it with results from PEAS. From our simulation results, we make three observations that show that our scheme perform better performance and match to analytical predictions. First, we assess our scheme and comparing it with PEAS algorithm. And Fig. 7 shows that our proposed sentinel scheme achieve better performance than PEAS [12]. The expected average energy spent falls considerably when our algorithm is compared to PEAS and we note that our algorithm enables lower energy consumption with a ratio of ' 36% of the total energy consumption. Second, we see beyond energy efficiency, that our solution permit to take into account the recurring nodes failure by dynamically adjusting nodes sleep timer to tend toward zero over time. Because nodes robustness fall over time and the probability of components failure become more important. And finally at our third observation, we see that our sentinel scheme support network scalability. In spite of all the computations are distributed in our scheme, Fig. 7 shows that growing the network density have not much more impact in the expected energy spent. The curves in the Fig. 6 show that the average energy consumption increases slightly with the number of network nodes. This is explained by the fact that the number of reserved nodes (nodes that probe their vicinity looking for a sentinel node) increases with the network density. And they often need to wake up themselves and check the presence of a sentinel node in the neighborhood, and these consumes some amount of energy due to probe messages exchanged.

### C.   Rapid maintenance evaluation

Our model is based on the probabilistic Weibull distribution to control sleeps nodes reserves. We applied a dynamic update of the Weibull scale parameter that is to say, the probing rate of nodes. Fig. 8 shows the evolution of the probing rate that increases as a function.

---

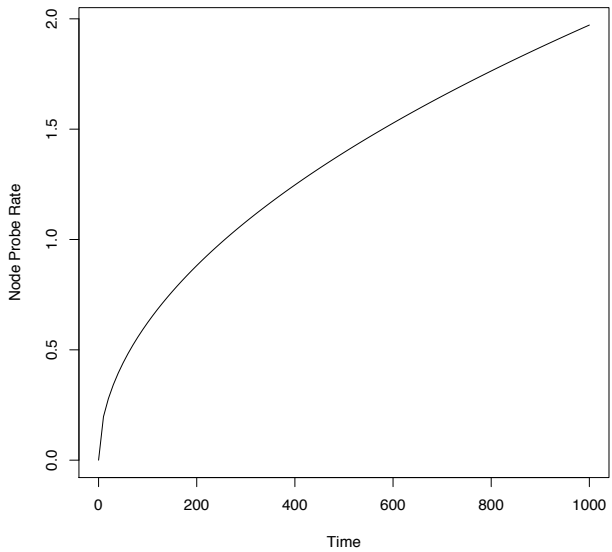

Figure 8. Node's probe rate adhustment over time

Once the probing rate obtained nodes use it to determine their sleep time. The standby time is inversely proportional to the probing rate (see figure 9). The nodes are decreasing their waking function of time and that in order to quickly replace a sentinel node that fails. Because, as we have raised above, the probability of failure of nodes increases over time.

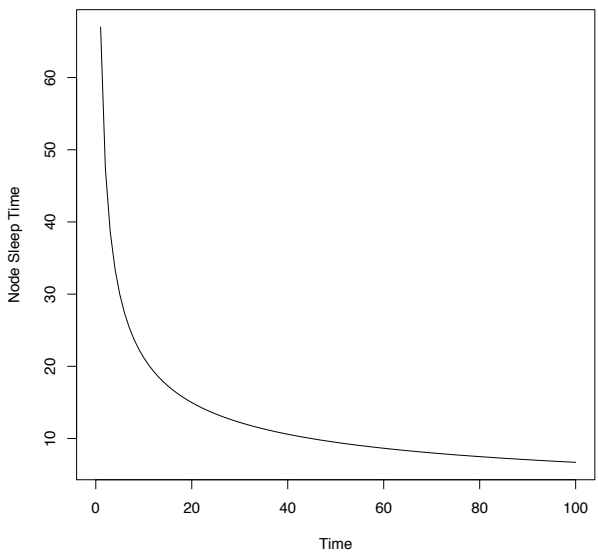

Figure 9. Node's sleep time updating over time

### D. Overhead Control

In our scheme, each node autonomously manages its sleep/wake up timer. And this is possible due to control messages exchanged during the probing step. This communication between nodes may generate flow overhead (Fig. 10 and Fig. 11) and then affect the network performances. Figure 10 shows that our scheme faced to the collision problem. And before, we ran samples simulation to assess collision impact through nodes communications. We first configure our simulation by ordering each probing node to sent one probe request to scan its vicinity and we see that after a few time over 60% of nodes passed to active mode. It mean that nodes does not receive probe request and hence consider that there are no sentinels. And quiet the same scenario is obtained when the number of sent probe request at each node is fixed to two messages per probing round. Fig. 11 shows that there is a big gap between the number of sent probe request and the received probe. This shows that most of the messages sent by probing nodes does not reach their destination i.e. the sentinel nodes. We observed that the probability that collisions occur growth proportionally with overhead i.e. with network's size. Comparing Figures 6 and 7, we note a proportionality between the number of received probes messages and the number of received probe responses that is to say that every one sentinel node that has received a probe message, it effectively sent a reply that came to the destination.That's why we fixed after several experiences, the number of probe request at each node to 3 messages and introduce the withdrawal algorithm to face the redundant active nodes problem due to collisions.

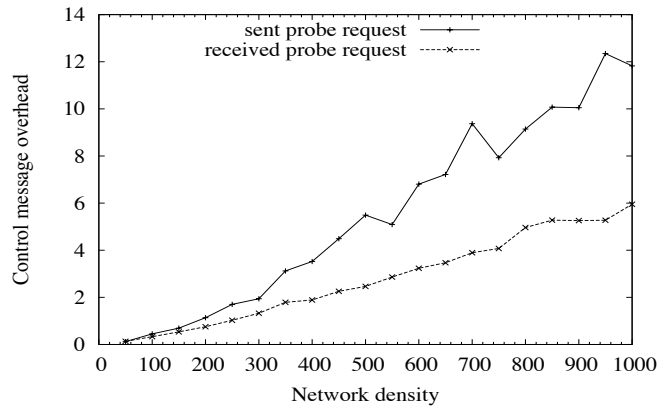

Figure 10. Control message overhead : sent probe requests vs. received probe requests

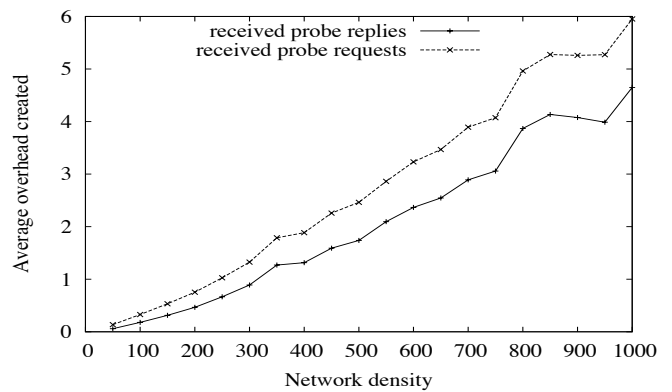

Figure 11. Control message overhead :received probe requests vs. received probe replies

## VI. CONCLUSION AND FUTURE WORKS

In this paper we analyze the design, implementation and experimental evaluation of a new scheduling algorithm based on sentinel scheme. The sentinel scheme exploit the redundancy offered by the cheap tiny sensor devices to ensure network accuracy and then prolong its lifetime. We also propose an energy aware sleep scheduling and rapid topology maintenance algorithm based on a sentinel scheme to enhance wireless sensor networks' lifetime. Our proposed scheme based on scheduling redundant nodes sleep periods, have several strengths. It permit first to schedule redundant nodes according to the Weibull distribution and guarantee an energy efficiency. And secondly, the Wiebull distribution helps to achieve autonomous operating nodes by dynamically adjusting node sleep time to take into account frequent nodes failure. Because, unlike in PEAS, our algorithm permit to dynamically adjust the nodes probe rate which is used to compute the sleep timer and no more need to keep into memory neighbors informations. Simulation results shows the robustness of our scheme by achieving energy efficient, scalable and fault tolerant algorithm. Through experimental figures, our proposed sentinel scheme presents better performances compare to PEAS.

Our future works include analyzing our scheme under the coverage problem and evaluate the lifetime evolution. And this will add more functionalities to our scheme and will make it more suitable for long life wireless sensor networks.

AUTHORS PROFILE

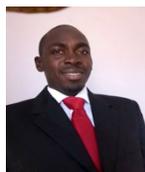

**Dame DIONGUE** is currently a Ph.D. student in computer science at LANI/Department of Computer Sciences of Gaston Berger University. He received a B. Sc. Degree in Computer Sciences from University Cheikh Anta Diop of Dakar, Senegal in 2007 and a M. Sc. Degree from University Cheikh Anta Diop of Dakar, Senegal in 2011. His areas of interest are wireless sensor networks, wireless mesh networks and optimization.

**Ousmane THIARE.** Received a PhD in computer science (Distributed systems) at 2007 from the university of Cergy Pontoise, France. He is an Associate Professor in Gaston Berger University of Saint-Louis Senegal. He has been author and co-author of published papers in several journals and recognized international conferences and symposiums.